\documentclass[pre,groupedaddress,showkeys,showpacs,twocolumn]{revtex4}
\usepackage{amsmath}
\usepackage{graphics}
\usepackage{graphicx}
\usepackage{amsfonts}
\usepackage{siunitx}
\usepackage{amssymb}
\usepackage{dcolumn}
\usepackage{bm}
\begin{document}
\title{Friction of poroelastic contacts with thin hydrogel films}

\author{Jessica Delavoipi\`ere}
\author{Emilie Verneuil}
\author{Yvette Tran}
\author{Antoine Chateauminois}
\email[]{antoine.chateauminois@espci.fr}
\affiliation{Soft Matter Science and Engineering Laboratory (SIMM), UMR CNRS 7615,Ecole Sup\'erieure de Physique et Chimie Industrielles (ESPCI), Universit\'e Pierre et Marie Curie, Paris (UPMC), France}
\author{Bertrand Heurtefeu}
\affiliation{Saint Gobain Recherche, Aubervilliers, France}
\author{Chung Yuen Hui}
\affiliation{Department of Mechanical and Aerospace Engineering, Cornell University, Ithaca, NY 14853, USA}
	\begin{abstract}
		We report on the frictional behaviour of thin poly(dimethylacrylamide) (PDMA) hydrogels films grafted on glass substrates in sliding contact with a glass spherical probe. Friction experiments are carried out at various velocities and applied normal loads with the contact fully immersed in water. In addition to friction force measurements, a novel optical set-up is designed to image the shape of the contact under steady-state sliding. The velocity-dependence of both friction force $F_t$ and contact shape is found to be controlled by a P\'eclet number Pe defined as the ratio of the time $\tau$ needed to drain the water out of the contact region to a contact time $a/v$, where $v$ is the sliding velocity and $a$ is the contact radius. When Pe<1, the equilibrium circular contact achieved under static normal indentation remains unchanged during sliding. Conversely, for Pe>1, a decrease in the contact area is observed together with the development of a contact asymmetry when the sliding velocity is increased. A maximum in $F_t$ is also observed at Pe~$\approx$~1. These experimental observations are discussed in the light of a poroelastic contact model based on a thin film approximation. This model indicates that the observed changes in contact geometry are due to the development of a pore pressure imbalance when Pe>1. An order of magnitude estimate of the friction force and its dependence on normal load and velocity is also provided under the assumption that most of the frictional energy is dissipated by poroelastic flow at the leading and trailing edges of the sliding contact.\\
	\end{abstract}
\maketitle
	%
	\section*{Introduction}
	\label{sec:introduction}
	Friction of hydrogel networks in aqueous solution pertains to many practical applications in the engineering and biomedical fields where they can often provide low friction. The lubricating properties of hydrogels are recognized to involve a complex interplay between the physical-chemistry of the surfaces, their ability to deform under hydrodynamic flow and stress-induced water transport phenomena within the gel network~\cite{gong2000a}. Among various physico-chemical aspects, experiments with polyelectrolyte gels have demonstrated friction can be affected by electrostatic repulsion  between surface charges or adsorption-desorption of polymer chains at the interface~\cite{kagata2001,oogaki2009,kamada2011,tominaga2008}. Another important phenomenon is the ability of the soft gel surface to deform under the action of hydrodynamical forces, thus allowing for the formation of a water lubricating film at the contact interface~\cite{jin1992,mccutchen1962,moore2014,smyth2017,skotheim2005}. Such elastohydrodynamic lubrication (EHL) effects are often reflected by the velocity-dependence of friction force which can exhibit a transition between the so-called boundary lubrication regime - where friction is mediated by physical interactions between contacting surfaces - and an elasto-hydrodynamic regime where the surfaces are progressively separated by a lubricant film as the velocity is increased~\cite{kurokawa2005,oogaki2009}. In combination with these elasto-hydrodynamic effects, the ability of the gel surfaces to support load without dehydrating has been shown to affect their frictional properties~\cite{dunn2013,reale2017}. The drainage of water-swollen polymer hydrogel under normal indentation by rigid probes has been widely evidenced and described~\cite{doi2009,chan2012b,galli2011,hu2010,hui2006,lin2007,strange2013,delavoipiere2016} within the framework of poroelastic theories~\cite{biot1955,doi2009,doi2008} which couple the elasticity of the gel network with pressure-induced water flow. During sliding, the contribution of such drainage phenomenon to friction can be seen from the dependence of frictional force on a P\'eclet number defined by the ratio of advective components (sliding) to diffusive components (fluid drainage)~\cite{moore2014,reale2017}. Interestingly, similar theories have been widely developed in the field of biotribology to describe articular cartilage lubrication as an alternative to classical elasto-hydrodynamic lubrications theories which often proved to be unable to describe experimental results~\cite{sakai2012,murakami2015,ateshian2009,yamaguchi2018}. Here, the lubricating properties are attributed to the ability of the pressurized pore fluid to support most of the load applied to the cartilage.\\
	In this study, we consider the contribution of poroelastic effects to the frictional behaviour of thin ($\mu$m) hydrogel layers mechanically confined between rigid glass substrates. As shown by previous normal indentation studies, the drainage of hydrogel layers under such conditions is indeed considerably enhanced as a result of the amplification of contact stresses arising from confinement~\cite{chan2012b,galli2011,hu2011}. Here, we address the relationship between poroelastic flow and sliding friction from experiments where we monitor both the friction force and the shape of the contact fully immersed in water as a function of velocity and normal force. Under contact conditions which are shown not to involve any substantial EHL effects, changes in the contact shape are found to be well predicted by a poroelastic contact model derived within the limits of a thin film approximation. Using this approach, we especially show that the progressive development of a pore pressure imbalance within the gel film generates a lift force which causes a reduction in contact size when the P\'eclet number is increased above unity. From this poroelastic contact model, we also derive an order of magnitude estimate of the friction force.\\ 
	In a first section, we detail the synthesis of the thin PDMA films and the custom-built frictional set-up. The experimental load- and velocity- dependence of steady-state friction of the films is then discussed in relation to \textit{in situ} contact vizualisation. A poroelastic model developed based on a thin film approximation is then derived which allows us to predict contact shape and contact stress from the pore pressure distribution and network elasticity. In the last section, experimental results are discussed in the light of this theoretical model.\\
	%
	%
	\section*{Experimental details}
	\label{sec:experimental}
	\subsubsection*{Synthesis of gel films}
	Poly(DMA) hydrogel films were synthesized by simultaneously crosslinking and grafting preformed polymer chains onto glass substrates with thiol-ene click reaction. As fully described earlier~\cite{cholletb2016,cholleta2016,li2015}, ene-reactive PDMA was first synthesized by free radical polymerization of dimethylacrylamide and acrylic acid which is then modified by amidification using allylamine. The ene-functionalized PDMA was subsequently purified by dialysis against water and recovered by freeze-drying.\\
	To ensure good adhesion between the hydrogel film and the glass substrate, we carried out thiol-modification of the borosilicate glass surfaces. Specifically, solid substrates (26x38x1~mm$^3$) were first cleaned in a freshly prepared 'piranha' (H$_2$SO$_4$/H$_2$O$_2$) solution, then rinsed and sonicated in Milli-Q water before drying under nitrogen flow. The freshly cleaned glass substrates were quickly transferred into a sealed reactor filled with nitrogen where a solution of dry toluene with 3 vol$\%$ of 3-mercaptopropyltrimethoxysilane was introduced. After 3 hours immersing in the silane solution under nitrogen, the glass substrates were rinsed and sonicated in toluene before drying.\\
	Surface-attached hydrogel films were synthesized by simultaneously crosslinking and grafting the ene-reactive preformed polymer by thiol-ene reaction. Prior to spin-coating, ene-functionalized PDMA and dithioerythritol crosslinker were dissolved in dimethylformamide (DMF). Spin-coating on the thiol-modified glass slides was performed for 30 seconds with angular velocity of 3000 rpm.  After spin-coating, polymer films were annealed at 150~$^\circ C$ for at least 16 hours under vacuum to activate the thiol-ene reaction. The glass substrates were then rinsed and sonicated in DMF and water to remove all free polymer chains, initiators and by-products from the polymer synthesis. This procedure resulted in PDMA films with thickness 1.65 $\mu$m in the dry state, as measured by ellipsometry. Ellipsometry measurements in water provided a swelling ratio of 1.9 for the fully hydrated films, i.e. a thickness of about 3.1~$\mu$m in the fully swollen state.\\
	\subsubsection*{Friction set-up}
\begin{figure}[!ht]
		\centering
		\includegraphics[width=0.6 \columnwidth]{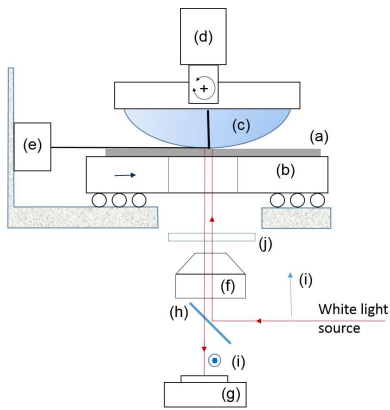}
		\caption{Schematic of the custom friction set-up. The coated glass substrate (a) fixed on a linear translation stage (b) is contacting a spherical glass probe (c) under an imposed normal force applied by a linear coil actuator (d) operated in closed-loop control. The friction force is measured along the contact plane using a strain gauge transducer (e) linked to the glass probe by steel strips. Images of the contact fully immersed in a droplet of water are recorded under white light illumination using a zoom lens (f), a CMOS camera (g) and an optical set-up consisting of a semi-reflecting plate (h), two crossed-polarizers (i) and a quarter-wave plate (j).}
		\label{fig:set_up}
	\end{figure}

	Friction experiments under imposed normal load conditions are carried out using the dedicated set-up schematically described in Figure~\ref{fig:set_up}. The coated glass is fixed to a translation stage and displaced laterally at imposed velocity (from 0.001 to 1~mm~s$^{-1}$) using a linear translation stage (M.403, PI, Germany). A linear coil actuator (V.275, PI, Germany) operated in force servo-control mode allows to apply a normal load in the range 50-1000~mN to a spherical glass lens with a radius of curvature $R=25.9$~mm. During friction experiments, the lateral force is continuously monitored using a strain gauge transducer (FN 3280, Measurements \& Specialities, France) attached to the glass lens assembly. Decoupling between the normal and lateral directions is achieved by means of a leaf springs arrangement. All experiments were carried out with the contact fully immersed within a droplet of deionized water. It was verified that capillary forces generated by the water meniscus are negligible with respect to the measured friction forces.\\
	During sliding, Reflection Interference Contrast Microscopy (RICM) images of the immersed contact were continuously recorded using a CMOS camera (600x600 pixels with 12 bits resolution) and the optical arrangement described in Figure~\ref{fig:set_up}. The use of a combination of crossed-polarizers and quarter-wave plates allowed us to visualize the immersed contact in spite of the low refractive index difference between the swollen gel film and the water. No damage of the films was seen from contact visualization. In addition, the prescribed contact conditions were selected in order that the water content $\phi$ of the gel network during sliding was always above the threshold corresponding to the onset of glass transition of PDMA (i.e. $\phi \approx 0.2$ as detailed in Delavoipi\`ere~\textit{et al}~\cite{delavoipiere2016}).\\
	%
	%
	\section*{Results}
	\label{sec:discussion}
\begin{figure}[!ht]
			\centering
			\includegraphics[width=0.7 \columnwidth]{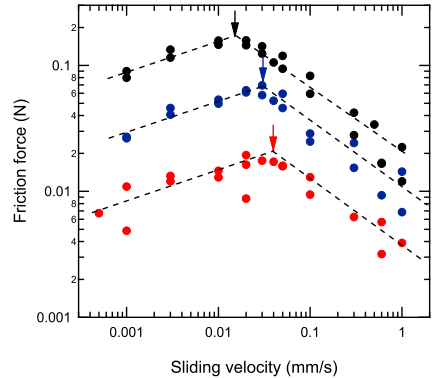}
			\caption{Friction force $F_t$ as a function of sliding velocity $v$ for various applied loads. Red: $F_n=$~50~mN, blue: $F_n=$~200~mN, black: $F_n=$~600~mN. The colour arrows indicates the location of the critical velocity $v_c$ at which the friction force is maximum. The dotted lines are guides for the eye.}
			\label{fig:ft_velocity}
		\end{figure}

\begin{figure}[!ht]
			\centering
			\includegraphics[width=0.5 \columnwidth]{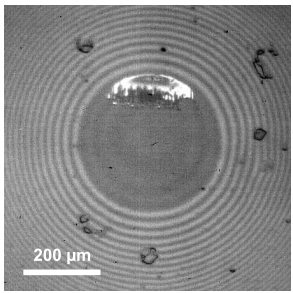}
			\caption{Contact image showing cavitation at the trailing edge ($F_n=1\,\si{N}$, $v=0.05\,\si{mm.s^{-1}}$). The coated glass substrate is moving from bottom to top with respect to the fixed glass lens.}
			\label{fig:contact_cavitation}
		\end{figure}

\begin{figure}[!ht]
		\centering
		\includegraphics[width=0.6 \columnwidth]{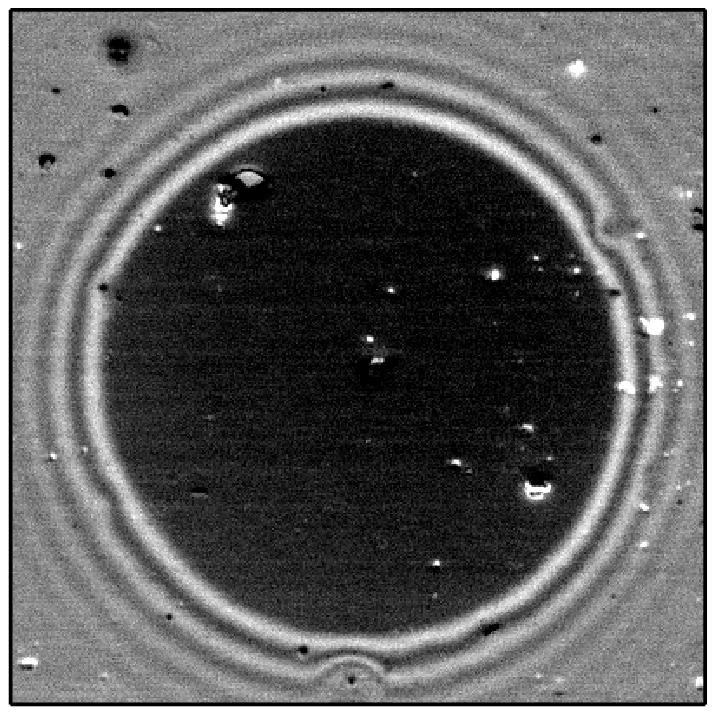}\\
		\includegraphics[width=0.6 \columnwidth]{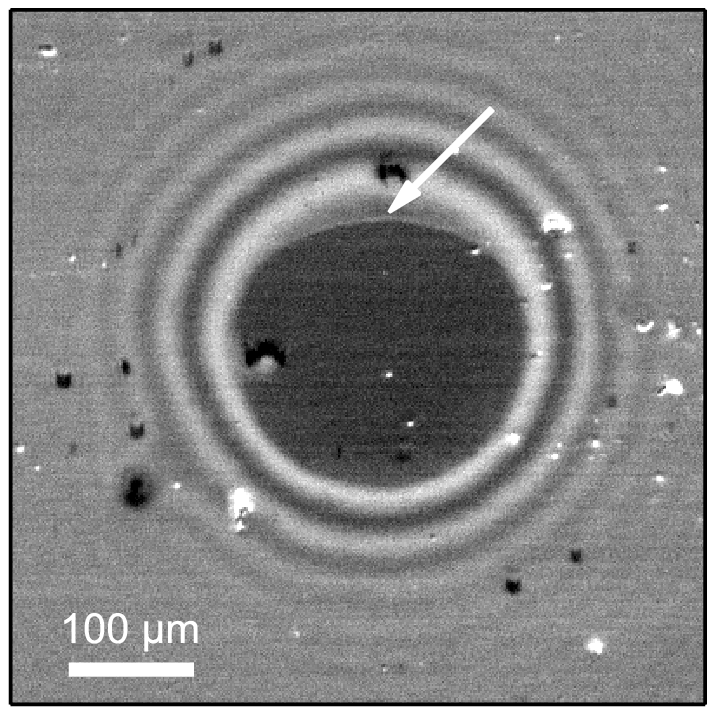}
		\caption{Contact images under steady state sliding ($F_n=600\,\si{mN}$). top: $v=10^{-3}\,\si{mm.s^{-1}}$; bottom: $v=1\,\si{mm.s^{-1}}$. The coated glass substrate is moving from bottom to top with respect to the fixed glass lens. As indicated by the white arrow, a loss of contact at the trailing edge is evidenced for $v=1\,\si{mm.s^{-1}}$. Rings are fringes of equal thickness in white light.}
		\label{fig:contact_pictures}
	\end{figure}

\begin{figure}[!ht]
		\centering
		\includegraphics[width=0.7 \columnwidth]{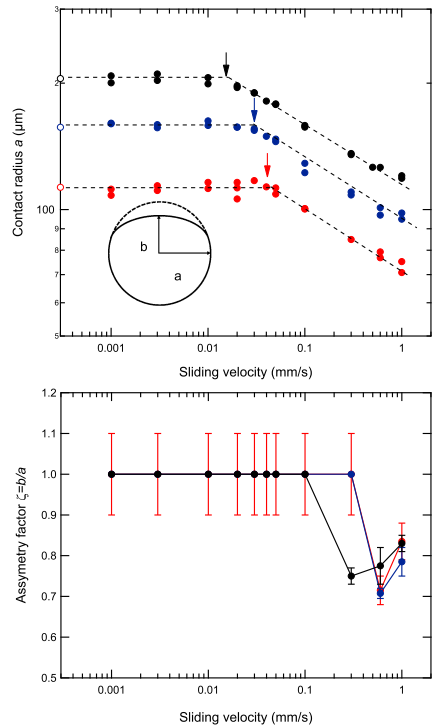}
		\caption{Contact shape as a function of sliding velocity. Top: contact radius $a$; bottom: asymmetry factor $\zeta=b/a$. Red: $F_n=$50~mN, blue: $F_n=$200~mN, black: $F_n=$600~mN. The colour arrows correspond to the characteristic velocity $v_c$ defined from the peak in friction force (see Figure~\ref{fig:ft_velocity}). In the top figure, open symbols on the left axis indicates the equilibrium contact radius measured under a static indentation load. Dotted lines are guides for the eye.}
		\label{fig:contact_shape}
	\end{figure}

\begin{figure}[!ht]
		\centering
		\includegraphics[width=0.7 \columnwidth]{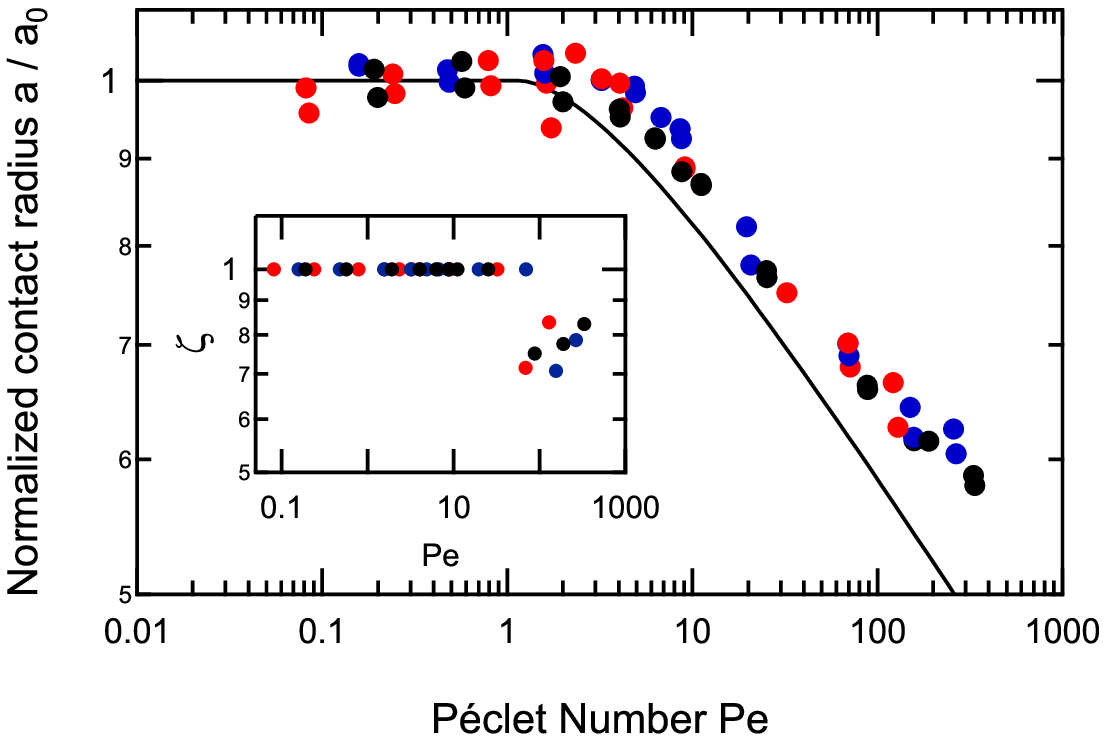}\\
		\includegraphics[width=0.7 \columnwidth]{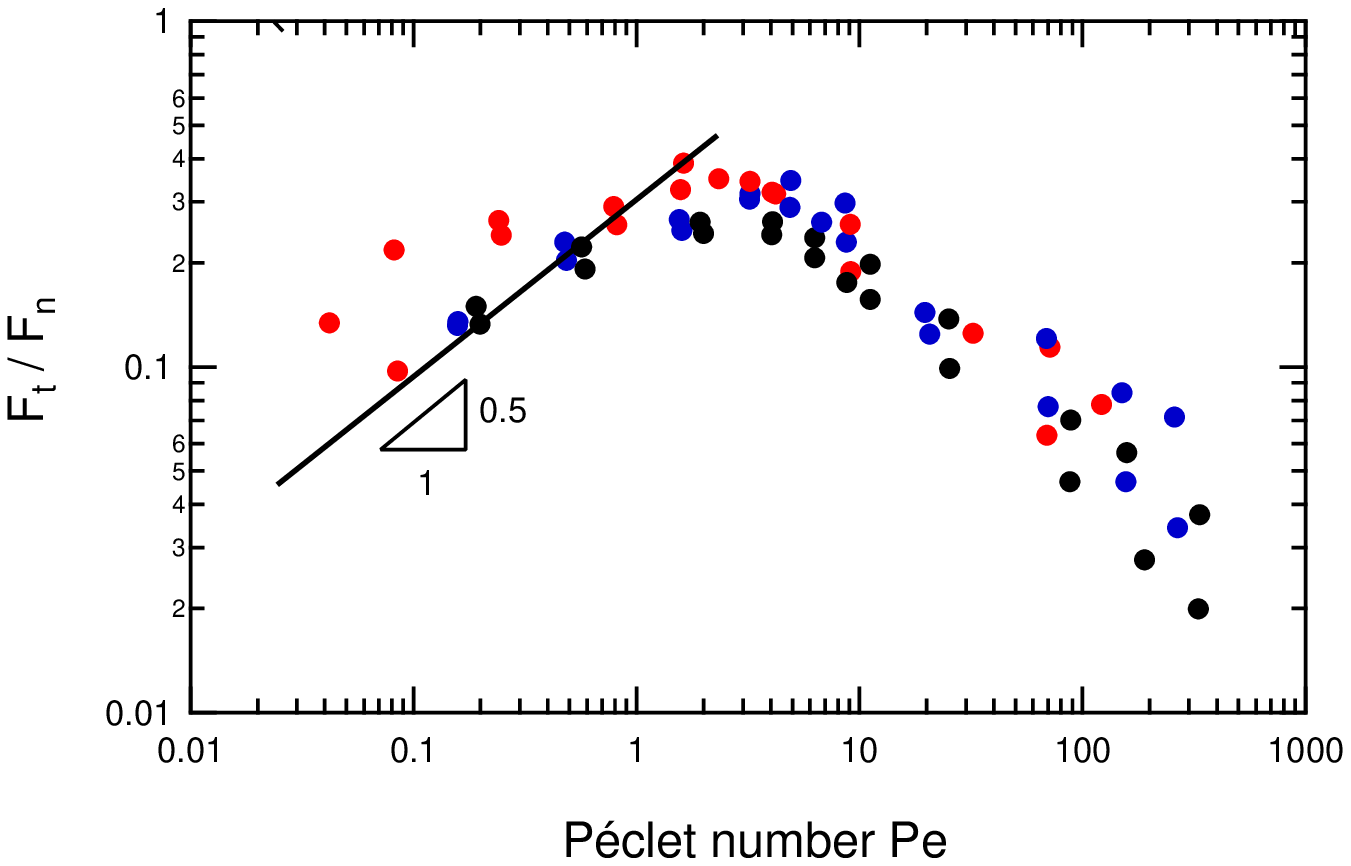}
		\caption{Normalized contact radius $a/a_0$ (top) and friction force $F_t/F_n$ (bottom) as a function of P\'eclet number Pe ($a_0$ and $F_n$ are respectively the equilibirum contact radius and the normal force). Inset: changes in the asymmetry parameter $\zeta$ as a function of Pe. The black line in the top figure corresponds to the prediction of the poroelastic model. In the bottom figure, the black line corresponds to a power law with exponent 0.5 as predicted by Eq.\ref{eq:friction:force}. Red: $F_n=$50~mN, blue: $F_n=$200~mN, black: $F_n=$600~mN.}
		\label{fig:a_Ft_Pe}
	\end{figure}

	Figure~\ref{fig:ft_velocity} details the velocity dependence of the steady-state friction force $F_t$ with different applied normal forces $F_n$ ranging from 50 to 600~$\,\si{mN}$. Within the investigated sliding velocity range, the friction force $F_t$ varies more than one order of magnitude. A friction force peak is observed at a critical velocity $v_c$ which slightly decreases when the applied normal load $F_n$ is increased ($v_c \approx $~0.015, 0.03 and 0.04~mm~s$^{-1}$ for $F_n=$~600, 200 and 50~mN, respectively). Below this critical velocity, the friction force increases with velocity according to a power law with exponent $0.25 \pm 0.05$ for all normal loads under consideration. Conversely, above the critical velocity, the friction force decreases with velocity according to another power law with exponent $-0.6 \pm 0.05$. At first sight, the observation of such a friction peak is reminiscent of the experimental velocity-dependence of dry rubber friction which is also characterized by a maximum in friction at some characteristic velocity (see e.g. Schallamach~\cite{schallamach1953}, Grosh~\cite{grosch1963} or Vorvolakos and Chaudhury~\cite{vorvolakos2003} among many others). This rubber friction peak has been attributed to either thermally activated pinning/depinning events of elastomer molecules to the contacting surface~\cite{schallamach1963} or to the occurrence of bulk viscoelastic effects at the (micro-)contact scale~\cite{grosch1963}. Although they cannot be completly discarded in our study, molecular pinning/depinning phenomena could not account for the changes in contact shape which are detailed below. Regarding viscoelastic effects, one expect they would be maximized when the characteristic contact frequency defined by $v/a$ (where $a$ is a characteristic contact size) is close to the glass transition frequency at the experimental temperature~\cite{grosch1963}. In our case, the characteristic frequency $v/a$ during sliding friction of the smooth spherical probe is less than 10~Hz, which is more than five orders of magnitude lower than the estimated glass transition frequency of the fully hydrated PDMA layer at room temperature. Therefore, any contribution of viscoelasticity to frictional energy dissipation can be neglected in our experiments.\\
	For sliding velocities higher than $1\,\si{mm.s^{-1}}$, stick-slip motions independent of the applied normal force were clearly indicated by the lateral force signal (not shown). In the corresponding velocity domain, the occurrence of such stick-slip motions is clearly favoured by the decreasing branch of the $F_t(v)$ friction law.\\
	When the normal load exceeds about 700~mN, cavitation is systematically observed at the rear of the contact (Figure~\ref{fig:contact_cavitation})  above a threshold velocity which is found to be about $0.1\,\si{mm.s^{-1}}$ and $0.05\,\si{mm.s^{-1}}$ for $F_n=700$~mN and $F_n=1$~N, respectively. This cavitation phenomenon is indicative of the occurrence of a negative pressure at the rear of the contact which will be discussed later in the light of our poroelastic contact model.\\
	Figure~\ref{fig:contact_pictures} shows two contact images taken below the cavitation threshold at $v=10^{-3}\,\si{mm.s^{-1}} < v_c$ and $v=1\,\si{mm.s^{-1}} > v_c$, respectively (the applied normal load is $F_n=$~600~mN and the critical steady-state velocity $v_c=0.015\,\si{mm.s^{-1}}$). A reduction in the contact size is clearly seen at $v=1\,\si{mm.s^{-1}}$. Moreover, while the circular shape of the initial static normal contact is preserved at low velocity, a contact asymmetry is present at $v=1\,\si{mm.s^{-1}}$. More precisely, a loss of contact is observed at the trailing edge while the leading edge retains its circular shape.\\
	The changes in contact shape as a function of normal load and sliding velocity are further detailed in Figure~\ref{fig:contact_shape}. As shown in the inset, the contact radii $a$ were determined from a measurement of the contact diameter perpendicular to the sliding direction. To describe the contact asymmetry, we define a  parameter $\zeta=b/a$ where $b$ and $a$ are contact radii along two crossed lines parallel ($b$) and perpendicular ($a$) to the sliding direction as depicted in Figure~\ref{fig:contact_shape} (top).
	Whatever the applied normal load, two distinct regimes are clearly distinguished depending on whether the velocity is lower or higher than the critical velocity $v_c$. In the low velocity regime (i.e. when $v < v_c$), the contact line remains circular ($\zeta \approx 1$) with a constant radius close to the radius, denoted $a_0$, achieved under static indentation loading at the same normal force (as indicated by the open symbols on the ordinate axis in Figure~\ref{fig:contact_shape} (top)). In the high velocity regime ($v > v_c$), the contact radius decreases progressively with increasing sliding velocity. At the same time, the asymmetry parameter $\zeta$ is first decreasing and then increasing.\\
	It therefore turns out that increasing the sliding velocity above a certain load-dependent threshold results in a concomitant decrease in both the contact area and in the friction force. Under a constant applied normal load, such a reduction in the size of the contact area is indicative of the development of a load-bearing capability within the contact. This phenomenon is reminiscent to what could be observed in fluid-lubricated contacts between a rigid sphere and a soft substrate. As described within the framework of elasto-hydrodynamic lubrication theories (see e.g. \cite{downson1977}), elastic deformation of the soft substrate  can couple normal and lateral forces and generates lift. This situation was theoretically addressed for a poroelastic layer by Sekimoto and Leibler~\cite{sekimoto1993} and Skotheim and Mahadevan~\cite{skotheim2005} who derived from scaling arguments an expression for the relationship between the lift force $F_l$ and a characteristic gap thickness $h_0$ defined in the absence of solid deformation. According to these approaches,
	\begin{equation}
	F_l \approx \frac{\eta^2v^2}{2\mu+\lambda}\frac{e_0R^2}{h_0^3}
	\end{equation}
	where $\eta$ is the fluid viscosity, $\mu$ is the shear modulus of the layer, $\lambda$ is a Lam\'e constant, $R$ is the radius of the spherical probe and $e_0$ is the thickness of the un-deformed layer. Using typical values of $\mu$ and $\lambda$ (see Appendix~A for an estimate of the mechanical properties of the PDMA film) and a gap thickness of $100$~nm, this expression yields a lift force on the order of 10$^{-7}$~N, i.e. 5 orders of magnitude lower than the applied normal force. This analysis suggests that elasto-hydrodynamic lubrication is negligible in our experiments and that any water film in between the two surfaces would be squeezed out of the contact. \textit{In situ} contact visualization also supports this conclusion as no gap between the gel film and the glass probe was observed under steady-state sliding.\\
	In what follows, we investigate an alternate description of the observed frictional response which is based on stress-induced water transport phenomena occurring within the gel network when the rigid slider is moved sideways. The relevance of such an approach can first be addressed by comparing the typical contact time $a/v$ to the characteristic time $\tau$ needed to drain water out the gel network under load. Accordingly, we define a P\'eclet number as the ratio of the diffusive to convection time
	\begin{equation}
	\textmd{Pe}=\frac{\tau v}{a}\: .
	\label{eq:peclet}
	\end{equation}
	As detailed in Appendix~A, the poroelastic time $\tau$ can be determined experimentally from separate normal indentation experiments. From the knowledge of $\tau$ and using the measured contact radius $a$, we determine the P\'eclet number for each sliding velocity and normal load.  In Figure~\ref{fig:a_Ft_Pe}, the contact radius is normalized by $a_0$, the contact radius under static normal indentation, and the friction force is normalized by the normal force $F_n$, and both are plotted against the P\'eclet number. Remarkably, both contact radius and friction force data collapse on single master curves. This shows that the velocity-dependence of $F_t$ and $a$ is dictated by the P\'eclet number. We also find that the condition Pe~$\approx$~1 corresponds to the transition between two regimes: for Pe<1, the contact radius is constant and equal to its equilibrium value under static conditions while it is decreasing with Pe above unity. Similarly, the maximum in friction force $F_t$ is achieved at Pe close to unity.
	This result motivates us to extend our previous static poroelastic contact model to sliding contacts~\cite{delavoipiere2016}.  Within the limits of geometrically confined contacts (thin film approximation), we derive expressions for the pore pressure and network stress which account for the generation of a lift force and the observed changes in the contact shape.\\
	%
	%
	\section*{Poroelastic contact model}
	\label{sec:model}
	We consider a rigid spherical probe with radius $R$ indenting a thin layer of hydrogel with initial thickness $e_0$. A normal force $F_n$ is imposed on the indenter. The indenter is then moved sideways in the horizontal direction $x$ at a steady velocity $v$. The contact edge is a curve defined by
	\begin{equation}
	r=a(\theta,v)\:\: \left | \theta \right | < \pi.
	\end{equation}
	Here $(r,\theta)$ is a polar coordinate system directly above the center of the rigid sphere. Note that the shape of the contact line in general depends on the Pe number. Because of symmetry, $a(\theta,v)$ is an even function of $\theta$ and we thus need to consider only the region $0 \leq \theta \leq \pi$ ($\theta$=0 along the $x>0$ axis).\\
	Our goal is first to determine the pore pressure field $p$ and the associated contact shape. As in Delavoipi\`ere \textit{et al}~\cite{delavoipiere2016}, we neglect the effect of shearing the elastic network. In fact, in the thin film approximation, the shear deformation is decoupled from the pore pressure field, hence it has no effect on the pressure. We assume steady state which means that all the fields are functions of $x=X-vt$ and $y$. Here $(x,y)$ is a moving coordinate system that is attached to the center of the contact point ($x=0$, $y=0$ at the center of contact) and $(X,y)$ is the position of a material point with respect to a fixed coordinate system. Thus,
	\begin{equation}
	\left. \frac{\partial }{\partial t}\right | _{X,y} \rightarrow -v \frac{\partial }{\partial x}
	\label{eq:coord}
	\end{equation}
	in steady state. Following Delavoipi\`ere~\textit{et al}~\cite{delavoipiere2016}, the thin film approximation tells us that
	\begin{equation}
	u_z=z\epsilon(r,\theta,v) \:,
	\end{equation}
	where
	\begin{equation}
	\epsilon (x)=\left\{\begin{matrix} 
	\left( \delta-r^2/2R \right)/e_0 & \left |r \right |<a(\theta,v)\\ 
	0  & \left |r \right |>a(\theta,v)
	\end{matrix}\right. 
	\label{eq:epsilon}
	\end{equation}
	is the normal strain in the film with $\delta$ the steady state indentation depth. According to the mixture theory developed by Biot~\cite{biot1955}, the stress is composed of two parts: one generated by the hydrostatic pressure of water within the pores and the other one induced by the contact stress within the polymer network. The normal contact stress $\sigma(x)$ is thus related to the strain by
	\begin{equation}
	\sigma(r,\theta)=\tilde{E} \epsilon (r,\theta) + p(r,\theta)
	\label{eq:sigma}
	\end{equation}
	with
	\begin{equation}
	\tilde{E}=\frac{2 \mu\left(1-\nu\right)}{1-2 \nu}
	\label{eq:oedometric}
	\end{equation}
	where $\mu$ is the shear modulus and $\nu$ is the Poisson's ratio of the drained network.\\
	The pore pressure field is governed by Darcy's law and steady state implies that
	\begin{equation}
	-v\frac{\partial \epsilon}{dx}=-\kappa \nabla^2p \: .
	\label{eq:steady_state1}
	\end{equation}
	where $\kappa=D_p/\eta$ with $D_p$ the permeability of the network and $\eta$ is the viscosity of the solvent.
	Next, in the $(r,\theta)$ coordinate system
	\begin{equation}
	\frac{\partial \epsilon}{\partial x}=\cos \theta \frac{\partial \epsilon}{\partial r} - \frac{\sin \theta}{r}\frac{\partial \epsilon }{\partial \theta}
	\label{eq:decomposition0}
	\end{equation}
	and
	\begin{equation}
	\nabla^2 p=\left[ \frac{1}{r}\frac{\partial}{\partial r} \left ( r\frac{\partial }{\partial r} \right )+\frac{1}{r^2} \frac{\partial^2}{\partial \theta^2}\right]p \: .
	\label{eq:decomposition1}
	\end{equation}
	Substituting Eq.~\ref{eq:epsilon} into Eq.~\ref{eq:steady_state1} and using Eqs.~\ref{eq:decomposition0} and \ref{eq:decomposition1}, we have
	\begin{equation}
	v \left ( \cos \theta \frac{\partial \epsilon}{\partial r} - \frac{\sin \theta}{r}\frac{\partial \epsilon }{\partial \theta} \right ) = \kappa \left [ \frac{1}{r}\frac{\partial}{\partial r} \left ( r\frac{\partial p}{\partial r} \right )+\frac{1}{r^2} \frac{\partial^2 p}{\partial \theta^2}\right ] \:.
	\label{eq:steady_state2}
	\end{equation}
	Inside the contact zone, the contact condition (\ref{eq:epsilon}) implies that $\epsilon$ is independent of $\theta$, Eq.~\ref{eq:steady_state1} reduces to
	\begin{equation}
	-r \cos \theta=\frac{Re_0\kappa}{v}\left [ \frac{1}{r}\frac{\partial}{\partial r} \left ( r\frac{\partial p}{\partial r} \right )+\frac{1}{r^2} \frac{\partial^2p}{\partial \theta^2}\right ] \:.
	\label{eq:pore_pressure}
	\end{equation}
	The general solution of Eq.~\ref{eq:pore_pressure} where $p$ is an even function of $\theta$ is
	\begin{equation}
	p=-\frac{v}{8Re_0\kappa}r^3  \cos \theta + \sum_{n=0}^{\infty} \alpha_nr^n \cos n \theta
	\label{eq:p_general}
	\end{equation}
	where $\alpha_n$'s are unknown coefficients to be determined from the following boundary condition
	\begin{equation}
	p(r=a(\theta,v))=0 \:
	\end{equation}
	where $a(\theta,v)$ is a function still to be determined.
	With $\delta=a^2/2R$, Eqs.~\ref{eq:epsilon} and \ref{eq:sigma} can be combined to derive the contact pressure
	\begin{equation}
	\sigma(r,\theta)=\frac{\tilde{E}}{2Re_0}\left(a^2-r^2 \right )+p(r,\theta)\: .
	\label{eq:sigma_general}
	\end{equation}
	%
	\subsection{Circular contacts}
	Let us first assume that the contact edge is a circle, that is $a(\theta,v) \equiv a_0$ independent of $\theta$ which is experimentally the case when Pe<1, so that 
	\begin{equation}
	\begin{split}
	p \left ( a_0,\theta \right )=0 \Rightarrow \left ( \alpha_1-\frac{v}{8Re_0\kappa}a_0^2 \right )a_0 \cos \theta\\
	+\sum_{n=0,n \neq 1}^{\infty} \alpha_n a_0^n \cos n\theta=0 \:.
	\end{split}
	\end{equation}
	The functions $\cos n\theta$ are mutually orthogonal, so this means that
	\begin{equation}
	p \left ( a_0,\theta \right )=0 \Rightarrow \alpha_1=\frac{v}{8Re_0\kappa}a_0^2 \:,\: \alpha_n=0,\:\:n=0,2,3,4,....
	\label{eq:a1_circ}
	\end{equation}
	Substituting Eq.~\ref{eq:a1_circ} into Eq.~\ref{eq:p_general} gives
	\begin{equation}
	p=\frac{v r}{8Re_0\kappa}\left ( a_0^2-r^2 \right )\cos \theta\:\:\:\:,\: r \leq a_0\:.
	\label{eq:p_circ}
	\end{equation}
	Next, we determine the contact pressure $\sigma(r,\theta)$. First, we substitute Eq.~\ref{eq:p_circ} into Eq.~\ref{eq:sigma_general} and enforce the contact condition that
	\begin{equation}
	\sigma(r,\theta)=\frac{\tilde{E}}{2Re_0}\left ( a_0^2-r^2 \right ) +\frac{vr}{8Re_0\kappa}\left ( a_0^2-r^2 \right )\cos \theta
	\label{eq:sigma_0}
	\end{equation}
	is positive inside the contact disk. If it is true for all $r < a_0$, then the circle assumption is valid. Of course, this needs not to be the case since $\cos \theta$ is negative for $\pi/2 < \theta \leq \pi$. Eq.~\ref{eq:sigma_0} becomes
	\begin{equation}
	\sigma(r,\theta)=\frac{\tilde{E}}{2Re_0}\left ( a_0^2-r^2 \right )\left [ 1+\frac{v}{4\tilde{E}\kappa} r \cos \theta \right ]
	\label{eq:sigma_0irc1}
	\end{equation}
	Therefore, as long as the sliding velocity is sufficiently low so that
	\begin{equation}
	\frac{v}{4 \tilde{E}\kappa}a_0<1 \:,
	\label{eq:velocity_cond}
	\end{equation}
	i.e. $v<v_c$ where
	\begin{equation}
	v_c=\frac{4 \tilde{E}\kappa}{a_0 }
	\label{eq:v_c}
	\end{equation}
	is a critical velocity, the entire circle $r<a_0$ is under compression and our assumption that $a=a_0$ holds.\\
	 According to Delavoipi\`ere \textit{et al}~\cite{delavoipiere2016}, the contact radius $a_0$ in this regime is related to the applied force $F_n$ by
		\begin{equation}
		a_0=\left ( \frac{4Re_0F_n}{\pi\tilde{E}} \right )^{1/4} 
		\label{eq:radius_equilibrium}
		\end{equation}
	which simply derives from the condition that integration of the stress over the contact area equals the normal force: $\int\int \sigma dS=F_n$.
	Furthermore, the poroelastic time $\tau$ can be expressed as (see appendix~A)
		\begin{equation}
		\tau=\frac{a_0^2}{4 \kappa \tilde{E}}\:.
		\label{eq:tau}
		\end{equation}	
	so that the above defined Peclet number (Eq.~\ref{eq:peclet}) writes:
	\begin{equation}
	Pe=\frac{v}{v_c}\frac{a_0}{a}\:,
	\label{eq:peclet:2}
	\end{equation}	
	Equation~\ref{eq:peclet:2} implies that the condition (\ref{eq:velocity_cond}) is equivalent to $\textmd{Pe}<1$. When Pe<1, the diffusion time is smaller than the convection time and the drainage state of the contact in this regime is the same as that achieved under a purely static loading under the same applied normal force. In this regime, the pressure(Eq.é\ref{eq:p_circ}) writes:
	
		\begin{equation}
		p=\frac{F_n}{2\pi a_0^2}\left ( 1-\frac{r^2}{a_0^2} \right )Pe \frac{r\cos \theta}{a_0}\:\:\:\:,\: r \leq a_0\:.
		\label{eq:p_circ:Pe}
		\end{equation}
	
	When Pe>1, Eq.~\ref{eq:velocity_cond} does not hold, the contact line shrinks non-uniformly and is no longer a circle. Physically, this corresponds to cases when the water flow through the network is too slow to relax the pressure unbalance within the typical contact time $a/v$.
	%
	%
	\subsection{Asymmetric contacts}
	We now consider the case where the sliding rate is sufficiently high so the contact line is no longer circular, that is
	\begin{equation}
	\frac{v}{v_c} > 1 \textmd{  or Pe>1   } 
	\end{equation}
	Then, the poroelastic flow is limited by the high convective rate such that the contact stress at the trailing edge of the contact becomes negative. First, Eq.~\ref{eq:sigma_0irc1} suggests that the normal stress is always compressive for $|\theta|\leq\pi/2$, therefore we make the hypothesis that the contact line at the leading edge remains a circular arc (perhaps with a different radius $a \neq a_0$).
We normalize the radial distance from origin $r$, the velocity $v$ and the pore pressure $p$ as follows
	\begin{subequations}
		\begin{align}
		\overline{r}\equiv & r/a_0 \\
		\overline{v}\equiv & v/v_c \\
		\overline{p}\equiv & p/ \left[ 2 F_n/\pi a_0^2 \right] 
		\end{align} 
	\end{subequations}	
	where $2 F_n/\pi a_0^2$ is the maximum value of the contact stress achieved at the center of the contact ($r=$0) under purely normal equilibrium conditions, i.e. when the pore pressure term vanishes. From Eq.~\ref{eq:radius_equilibrium}, $2 F_n/\pi a_0^2=\tilde{E}a_0^2/(2Re_0)$. In terms of these normalized variables, the pressure given by Eq.~(\ref{eq:p_general}) is 
	\begin{equation}
	\overline{p}=-\overline{r}^3\overline{v}\cos\theta+\sum_{n=0}^{\infty}\overline{\alpha}_n\overline{r}^n\cos n\theta
	\label{eq:pressure_adim}
	\end{equation}
	with
	\begin{equation}
	\overline{\alpha_n}=\frac{\alpha_na_0^n}{\frac{2F}{\pi a_0^2}}
	\end{equation}
	From Eqs.~\ref{eq:sigma_general} and \ref{eq:pressure_adim} and using for $\sigma$ the same normalization as for $p$, the normalized contact stress is
	
	\begin{equation}
	{\overline{\sigma}}= \left ( \overline{a}^2- \overline{r}^2 \right ) -\overline{r}^3 \overline{v}\cos\theta+\sum_{n=0}^{\infty}\overline{\alpha}_n\overline{r}^n\cos n\theta \:,
	\label{eq:contactstress_adim}
	\end{equation}
	where	
	\begin{equation}
	\overline{a} \equiv a/a_0 
	\end{equation}
	is the normalized radius of the front contact line that remains circular (an unknown unless we know the indentation depth $\delta$ from experiment). As detailed in Appendix~B, Eq.~\ref{eq:contactstress_adim} can be solved numerically to determine the contact line and the associated pore pressure and contact stress distributions for Pe number greater than unity. In the section below, we discuss the experimental results in the light of these poroelastic contact calculations.
	%
	%
	\section*{Discussion}
\begin{figure}[!ht]
		\centering
		\includegraphics[width=0.8 \columnwidth]{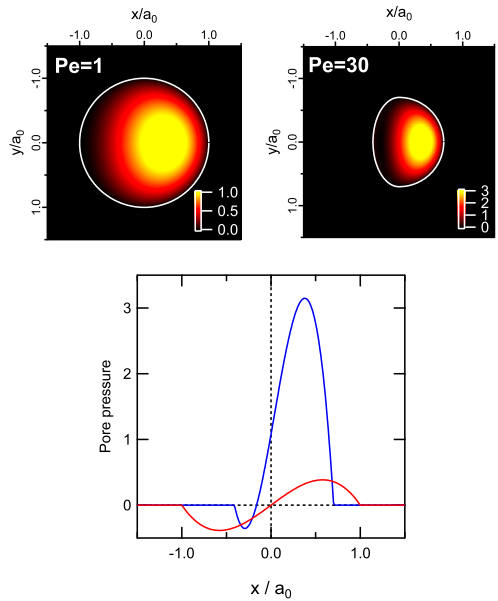}
		\caption{Poroelastic contact calculations for Pe$\geq$1. Top: normalized contact pressure distributions for Pe=1 and Pe=30. The white lines delimit the calculated contact area. The spherical probe is moving from left to right. Bottom: normalized pore pressure profiles taken along the sliding direction in the contact mid-plane for Pe=1 (red) and Pe=30 (blue).}
		\label{fig:contact_simulations}
	\end{figure}

	When Pe<1, experiments and theory agrees well: a circular contact shape is achieved with a radius corresponding to the equilibrium poroelastic state achieved under normal indentation. When the P\'eclet number is increased above unity, experimental observations under steady-state sliding show both a decrease in contact size and the progressive development of a contact asymmetry. The relevance of poroelastic flow to explain this phenomenon is supported by the theoretical simulations in Fig.~\ref{fig:contact_simulations}. The top part of the figure shows the normalized normal stress distributions determined for Pe=1 and Pe=30 with the white lines corresponding to the calculated contact line. As expected, the contact is circular with $a/a_0=1$ for Pe=1. Conversely, the predicted contact shape for Pe=30 is clearly asymmetric with $a/a_0<1$. Pore pressure profiles taken along the sliding direction and across the contact are detailed in the bottom part of the figure. When Pe=1, the profile is symmetrical and the net contribution of pore pressure to the normal force is thus zero. On the other hand, at Pe=30, the asymmetry in the pore pressure profile results in the generation of a net lift force which accounts for the reduced contact size as Pe is increased. The magnitude of these effects is enhanced at high Pe as both the amplitude of the pore pressure distribution and its asymmetry are increasing with Pe. The poroelastic contact model thus provides a consistent description of the trends in contact shape when Pe is varying above unity.\\ 
	Figure~\ref{fig:a_Ft_Pe} (top) displays a quantitative comparison between experimental and theoretical data for the relative change in contact radius $a/a_0$ as a function of Pe number. It turns out that the model provides a satisfactory, although slightly overestimated, prediction of the experimental reduction in contact radius for Pe>1. The resolution in the measurement of the asymmetry parameter $\zeta$ as a function of Pe does not allow for a detailed comparison with the model. However, the minimum measured values in $\zeta$ (about 0.7) are consistent with the prediction of the model ($\zeta=0.56$ for Pe=50).\\ 
	The close agreement between theory and experiments thus strongly suggests that the geometry of the contact line during sliding can be well explained by our poroelastic model. However, it should be noted that our model does not predict the increase in $\zeta$ which is observed at the highest velocities (see Figure~\ref{fig:contact_shape} (bottom)). This is due to the limitation of our foundation model, which must break down in the high velocity regime where the drainage of the layer vanishes. Indeed, the layer becomes incompressible at high sliding velocities and the resulting deformation cannot be captured by our one-dimensional foundation model. In this purely elastic regime, the contact shape is expected to become circular again (in the absence of any shear effects). If shear stresses are neglected, contact radius in this regime can be computed from an exact elastic contact mechanics model for layered substrates developed by Perriot \textit{et al}~\cite{perriot2004}. A calculation using this model indicates that the relative change in contact radius $a/a_0$ for the incompressible film is close to 0.60, 0.55 and 0.50 for $F_n=$~50, 200 and 600~mN, respectively. These values for an incompressible layer are lower than that measured at the highest Pe numbers and can thus be viewed as lower bound limits for contact size reduction.\\
	The foundation model is also limited in that the effects of shear are neglected, so it cannot be used to determine the friction force directly. In contact mechanics, the trailing and leading edges of the sliding indenter can be viewed as steady state growing cracks~\cite{baumberger2002,Johnson1997a,Savkoor1992}. Specifically, the trailing edge is an advancing crack while the crack tip at the leading edge is receding. In our case these cracks are growing/receding under mixed mode conditions (predominant Mode II). The energy needed to drive these sliding cracks has at least two contributions: the energies dissipated by poroelastic flow and viscoelasticity of the layer. As mentioned earlier, there is little or no viscoelasticity in our system at room temperature, so most of the energy dissipated (per unit crack area) is by poroelastic flow. This energy can be estimated using our foundation model in the regime where Pe$\leq$~1 and is found to be (see Appendix~C)
	\begin{equation}
	\mathcal{G}_{poro}\equiv \frac{1}{6\pi}\frac{F}{R}Pe
	\label{eq:Gporo}
	\end{equation}
	For Pe>1, the energy dissipated can be found numerically but the order of magnitude is the same. The value of $\kappa$ can be estimated from independent indentation experiments (see Appendix~A). It is found to be about $\kappa=$~5~10$^{-18}$~N$^{-1}$~m$^4$~s$^{-1}$. Based on these estimates and using Eq.~\ref{eq:Gporo}, the energy dissipated per unit crack area by poroelastic flow is found to be on the order of 1~J~m$^{-1}$.  Note that Eq.~\ref{eq:Gporo} is an upper estimate since it assumes that all the energy is dissipated at the crack tip which is not necessarily the case.\\
	Then, the friction stress $\sigma_f$ can be estimated using a fracture mechanics approach where the applied energy release rate $\mathcal{G}_{app}$ must be equal to the energy needed to grow the crack per unit area. Without doing a full 3D analysis, it is not possible for us to determine the applied energy release rate. Here we used a scaling argument to estimate $\mathcal{G}_{app}$. The friction stress can be considered as an external shear load on the layer. A simple scaling argument shows that $\mathcal{G}_{app}$ can be written as
	\begin{equation}
	\mathcal{G}_{app} \approx \frac{\sigma_f^2}{E^*}e_0 \:\:\: \mathrm{      ,   Pe} \leq 1
	\label{eq:Gapp}
	\end{equation}
	where $E^*=E/(1-\nu^2)$ is the plane strain modulus of the layer. In our system, the Poisson's ratio varies between 0.45 (fully drained) and 0.5. Since we are interested in a scaling argument we used  $E^*=\tilde{E}(1-2\nu)/(1-\nu)^2 \approx \tilde{E}/3 $.  Using the energy balance condition $\mathcal{G}_{app}=\mathcal{G}_{poro}$, the shear stress is found to be:
	\begin{equation}
	\sigma_f \approx \left( \frac{1}{18\pi}\frac{\tilde{E}F_n}{R e_0}\right)^{1/2}\sqrt{Pe} \:\:\: \mathrm{   , Pe} \leq 1,
	\label{eq:shear_stress_Pe}
	\end{equation}
	The friction force simply writes $F_t=\sigma_f\pi a_0^2$ so that:
	\begin{equation}
		F_t \approx \frac{\sqrt{2}}{3}F_n\sqrt{Pe} \:\:\: \mathrm{   , Pe} \leq 1,
		\label{eq:friction:force}
		\end{equation}
 	First, Equation~\ref{eq:friction:force} predicts the collapse of all friction force data when normalized by the normal force $F_n$ and plotted against Pe, in remarkable agreement with our experimental data in Fig.\ref{fig:a_Ft_Pe}. Second, Eq.~\ref{eq:friction:force} provides an estimate of the normalized friction force: $F_t/F_n \sim 0.5$ for $\mathrm{Pe}=1$.
	 This prediction compares well with the experimental results shown in Fig.~\ref{fig:a_Ft_Pe}-bottom given the scaling approach we used. Finally, Eq.~\ref{eq:friction:force} predicts a variation of $F_t/F_n$ with Pe with an exponent $1/2$. Although we should be cautious about the fact that available data only span over 1.5 decade in P\'eclet, we find that a power law with exponent $1/2$ fits the data reasonably well. The discrepancy at low applied load ($F_n=50~mN$, red dots) could reflect the occurrence of additional dissipative processes at the glass/gel interface which are not taken into account in our fracture mechanics model.\\
	The calculation of applied energy release rate becomes much more complicated when Pe>1. Indeed, the contact area being no longer symmetric, the applied energy release rate needs to include the contribution due to the unsymmetrical normal pressure distribution. In this case, the applied energy release rate is taken to be the sum of the shear energy release rate given by Eq.~\ref{eq:Gapp} and a contribution from the normal indentation, i.e.,
	\begin{equation}
	\mathcal{G}_{app} \approx \frac{\sigma_f^2}{E^*}e_0+\alpha(\zeta) \frac{\sigma_n^2}{E^*}e_0
	\label{eq:Gapp_Pe_gt_1}
	\end{equation}
	where $\alpha(\zeta)$ is a positive scaling constant associated with the asymmetric shape factor $\zeta$ and is required to vanish at $\zeta=1$  which allows us to recover Eq.~\ref{eq:Gapp} when Pe<1. Physically, we expect $\alpha(\zeta)$ to be decreasing function of $\zeta$ (it reaches 0 at $\zeta=1$). Note that since $\sigma_n>>\sigma_f$ in our experiments, the second term in equation~\ref{eq:Gapp_Pe_gt_1} can significantly reduce the friction stress even if $\alpha(\zeta)$ is small. This is qualitatively consistent with the experimental observation that the frictional stress is decreasing with increasing velocity or Pe when Pe>1.
	\section*{Conclusion}%
	In this study, we have investigated the steady-state friction of thin hydrogel layers mechanically confined within contacts between rigid glass substrates. The role of stress-induced poroelastic flow within the gel network was reflected by the dependence of the frictional force and contact size on the P\'eclet number defined by the ratio of advective components (sliding) to diffusive components (fluid drainage). In addition, contact imaging provided new insights into associated changes in contact shape. While the equilibrium contact shape achieved under static indentation loading remains unchanged under steady-state sliding when Pe<1, a reduction in contact radius and the development of contact asymmetry were evidenced for Pe>1. Using a poroelastic contact model based on a thin film approximation, we show that these changes in the contact shape with Pe can be quantitatively described. The main ingredient of the model is the progressive development in a pressure imbalance within the gel network between the leading and trailing edges of the contact which results in the generation of a lift force. These poroelastic phenomena present some similarities with EHL lubrication where a thin fluid-film is generated at the contact interface. In this later case, a load-carrying capacity is also generated by an imbalance in the fluid film pressure distribution as a result of the coupling between the elasticity of the substrates and hydrodynamic forces.\\
	Our model also reveals that poroelasticity can account for a significant part of the measured friction force. We demonstrate that most of the frictional dissipation can be accounted for by a fracture mechanics approach where the leading and trailing edges of the contact are viewed as closing and opening cracks, respectively, where dissipation arises from poroelastic flow. Naturally, this does not preclude the occurrence of other dissipative processes related to physico-chemical interactions at the sliding interface. Such phenomena could also be dependent on P\'eclet number as the density of surface chains available for bonding will also be affected by the extend of network drainage. Such effects would deserve further studies where, for example, the physical-chemistry of the surface is modified.\\
	From the application view point, our theoretical model can provide some guidelines on how network properties (modulus, permeability) and film thickness should be tuned to tailor specific frictional properties. One way of doing this is to adjust the Pe number for tribological conditions encountered in a specific application. For thin films, we showed that this task is indeed facilitated by the explicit relationship between Pe and the elastic and permeation properties of the gel network.\\
	
	\appendix
	\section{Determination of the poroelastic time $\tau$ from normal indentation experiments}
	\label{sec:SI1}
\begin{figure}
		\centering
		\includegraphics[width=0.7 \columnwidth]{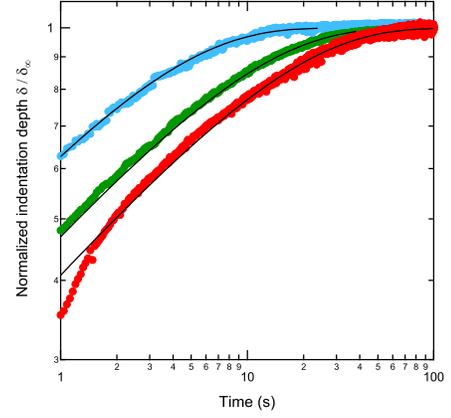}
		\caption{Log-log plot of the measured normalized indentation depth $\delta/\delta_{\infty}$ versus time for three applied normal loads. Red: $F_n=$~50 mN ; green: $F_n=$200 mN ; blue: $F_n=$600 mN. Black solid lines correspond to fits of the experimental data to Eq.~\ref{eq:poro_indent}}.
		\label{fig:SI1_fig1}
	\end{figure}

		\renewcommand{\theequation}{SI1.\arabic{equation}}
	To determine the elastic and permeation properties of the PDMA films, indentation experiments are carried out under a constant applied normal force $F_n$ with the contact loaded at $t=0$ within a short time. The results are analysed using a previously developed poroelastic contact model which was derived within the limits of thin film approximation with rigid substrates (Delavoipi\`ere \textit{et al} \textit{Soft Matter} (2016) \textbf{12} 8049-8058). For the sake of clarity, we briefly summarize the main results. The indentation depth $\delta$ is related to the indentation time $t$ by 
	\begin{equation}
	\frac{t}{\tau}=-\frac{\delta}{\delta_{\infty}}+\frac{1}{2}log\left(\frac{1+\delta/\delta_{\infty}}{1-\delta/\delta_{\infty}}\right)
	\label{eq:poro_indent}
	\end{equation}
	where $\delta_{\infty}$ is the equilibrium value of the indentation depth and $\tau$ is the characteristic poroelastic time defined by
	\begin{equation}
	\tau=\frac{1}{2 \kappa}\left[\frac{Re_0F_n }{\pi \tilde{E}^3}\right ]^{1/2} \:,
	\label{eq:tau}
	\end{equation}
	with $R$ the radius of the spherical probe, $e_0$ the thickness of the swollen film and $\tilde{E}$ the uniaxial compression modulus. The equilibrium value of the indentation depth $\delta_{\infty}$ depends on contact parameters ($R$,$F_n$) and on film properties ($e_0$, $\tilde{E}$):
	\begin{equation}
	\delta_{\infty}=\left[ \frac{F_ne_0}{\pi R \tilde{E}}\right]^{1/2} \: .
	\label{eq:delta_inf}
	\end{equation}
	The contact radius $a$ is related to indentation depth $\delta$ by the following geometrical relationship 
	\begin{equation}
	\delta=\frac{a^2}{2R} \:.
	\label{eq:delta}
	\end{equation}
	Accordingly, the equilibrium contact radius $a_0$ is
	\begin{equation}
	a_{0}=\sqrt{2 R \delta_{\infty}}=\sqrt{2}\left[\frac{F_ne_0R}{\pi \tilde{E}} \right ]^{1/4}
	\label{eq:a_infty}
	\end{equation}
	Using Eq.~\ref{eq:a_infty}, Eq.~\ref{eq:tau} can be rewritten as
	\begin{equation}
	\tau=\frac{a_{0}^2}{4 \kappa \tilde{E}}
	\end{equation}
	During experiments, the contact radius $a$ is continuously monitored as a function of time under a constant applied normal load and translated into indentation depth $\delta$ using Eq.~\ref{eq:delta}. Then, a fit of the experimental data to Eq.~\ref{eq:poro_indent} (see Fig.~\ref{fig:SI1_fig1}) provides the poroelastic time $\tau$. For the PDMA film under consideration, fitted values of $\tau$ are found to be 9.1, 25.3 and 39.8~s for $F_n$=50, 200 and 600~mN, respectively.\\
	Using Eq.~\ref{eq:a_infty} and the measured equilibrium value of the contact radius $a_0$, the oedometric modulus can be estimated to be $\tilde{E}=43 \pm 4$~MPa.\\
	Then, an estimate of $\kappa$ can be deduced from Eqs.~(\ref{eq:tau}) using the experimental values of $\tau$ and $\tilde{E}$. A value of $\kappa$ in the range 5-6~10$^{-18}$~m$^2$~Pa$^{-1}$~s$^{-1}$ is obtained. Taking $\nu=0.45$ for the Poisson's ratio of the drained network, one gets a value of $\mu \approx 1$~MPa and $\lambda \approx 10$~MPa for the shear modulus and the Lamé constant, respectively.
	These values are relatively high for a gel network. However, it should be pointed out that, for the purpose of friction study, we had to achieve a relatively high level of cross-link density within the gel network otherwise films are damaged during sliding experiments. The increased crosslink density of the network is evidenced by the swelling ratio of the investigated PDMA film which was only 1.9 instead of the value of 3 that we used in a previous indentation study.$^{21}$
	\section{Numerical implementation of the poroelastic sliding model for Pe$>$1}
	\label{sec:SI2}
\begin{figure}
	\centering
	\includegraphics[width=0.7\linewidth]{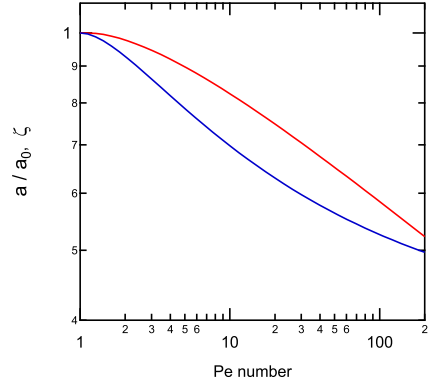}
	\caption{Calculated relative changes in contact radius $a/a_0$ (red) and in asymmetry parameter $\zeta$ (blue) as a function of Pe number. $a_0$ is the equilibrium contact radius achieved for Pe$\leq$1.}
	\label{fig:SI2_Fig1}
\end{figure}

\begin{figure}
	\centering
	\includegraphics[width=0.7\linewidth]{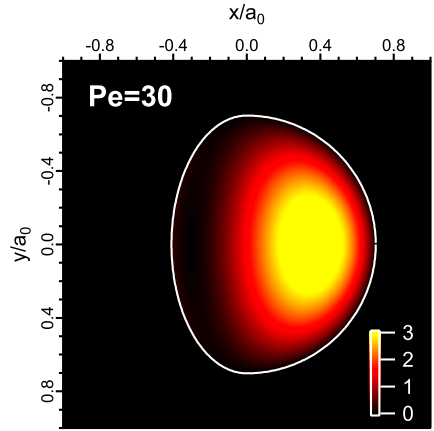}
	\caption{Calculated normalized contact stress distribution for $\mathrm{Pe}=30$. The white line represents the contact line where $\overline{p}=0$.}
	\label{fig:SI2_Fig2}
\end{figure}

\begin{figure}
	\centering
	\includegraphics[width=0.7\linewidth]{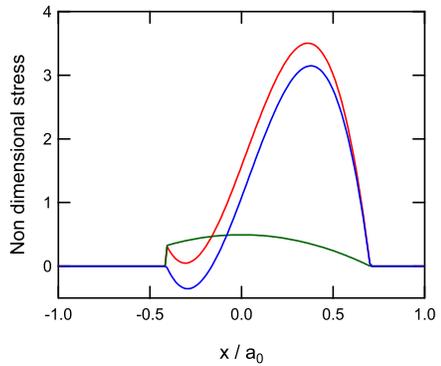}
	\caption{Profiles of calculated normalized contact stress $\overline{\sigma}$ (red), pore pressure $\overline{p}$ (blue) and network stress $\overline{\sigma}=\overline{a}^2-\overline{r}^2$ (green) along $x$ direction ($y=0$) for $\mathrm{Pe}=30$.}
	\label{fig:SI2_Fig3}
\end{figure}

	\renewcommand{\theequation}{SI2.\arabic{equation}}
For Pe>1, a numerical solution for the contact line and the associated pore pressure was obtained from Eq.~30 assuming that the contact line has the general form $\overline{r}=f(\theta)$ where $f$ is an even function of $\theta$ 
\begin{equation}
f(|\theta| \leq \pi/2)=\overline{a}
\end{equation}
Consistently with the experimental observations, the curve $f$ is assumed to be described by an ellipse for $|\theta| \geq\pi/2$
\begin{equation}
\begin{split}
\overline{x}=&\zeta \overline{a} \cos \theta\:\:\: |\theta| \geq\pi/2 \\
\overline{y}=&\overline{a} \sin \theta
\end{split}
\end{equation}
where $0<\zeta<1$ is a numerical parameter describing the contact asymmetry. For each velocity $v>v_c$, the two unknowns $\overline{a}$ and $\zeta$ are determined from an iterative procedure using Eqs.~30 and 32 for the pore pressure and contact stress, respectively.\\
We increase $\overline{v}$ from 1 to $1+\Delta \overline{v}$ where $\Delta\overline{v}$ is a small positive number. Then, the condition $\overline{\sigma}>0$ is no longer satisfied by setting $\overline{\alpha}_1=1$, $\overline{\alpha}_0=0$, $\overline{\alpha}_2=\overline{\alpha}_3=...=0$ in Eq.~32 inside the circle $\overline{a}=1$. Indeed, if we set $\overline{\alpha}_1=1$, $\overline{\alpha}_0=0$, $\overline{\alpha}_2=\overline{\alpha}_3=...=0$ when $\overline{v}=1+\Delta \overline{v}$, the pressure given by Eq.~30 is no longer vanishing on the circle $\overline{a}=1$ and the boundary where $\overline{\sigma}=0$ is given by
\begin{equation}
0=\overline{r} \cos \theta \left( 1- \overline{r}^2 \overline{v}  \right)+\left(1- \overline{r}^2\right)
\end{equation}
which defines a new contact line.\\
Accordingly, we decrease $\zeta$ from 1 to $1-\Delta\zeta$ where $\Delta\zeta$ is a small positive number to obtain a new contact line. From Eq.~30, we determine the set of $\overline{\alpha}_n$ parameters ensuring $\overline{p}=0$ on this newly defined contact line. Using this new set of $\overline{\alpha}_n$ parameters, the contact stress is calculated from Eq.~32. Then, we calculate the resulting normal force from numerical integration of the contact stress (Eq.~32):
\begin{equation}
\begin{split}
&\int \int _A \sigma dA=F_n
\end{split}
\end{equation} 
where $A$ is the contact area defined by $\zeta=1-\Delta\zeta$ and $\overline{a}=1$. If the condition $F_n=F_{imp}$ is violated ($F_{imp}$ is the imposed force), then we have to reduce $\overline{a}$ from 1 to $1-\Delta\overline{a}$ where $\Delta\overline{a}$ is a small positive number. This defines a new $\zeta'$ parameter
\begin{equation}
\zeta'=\frac{1-\Delta\zeta}{1-\Delta\overline{a}}
\end{equation}
Then we calculate again the set of $\overline{\alpha}_n$ parameters ensuring $\overline{p}=0$ on the updated contact line and keep iterating until both conditions
\begin{equation}
\begin{split}
&\overline{p}=0 \:\: \text{on the contact line}\: f(\theta)\\
&\int \int _A \sigma dA=F_n
\end{split}
\end{equation} 
are satisfied.\\
The calculated values of $\overline{a}=a/a_0$ and $\zeta$ are reported as a function of Pe in Fig.~\ref{fig:SI2_Fig1}. Figure~\ref{fig:SI2_Fig2} shows an example of the calculated contact stress distribution within the contact for Pe=30. Profiles of contact stress $\overline{\sigma}$, pore pressure $\overline{p}$ and network stress $\overline{\sigma}=\overline{a}^2-\overline{r}^2$ are also detailed in Fig.~\ref{fig:SI2_Fig3}.
	\section{Estimate of the energy dissipated by poroelastic flow under steady state sliding}
	\label{sec:SI3}
	\renewcommand{\theequation}{SI3.\arabic{equation}}
We make the assumption that all the dissipated energy during sliding friction is associated with crack closure and opening mechanisms at the leading and trailing edges of the contact, respectively. We also assumed that energy dissipation at the crack tip is solely arising from poroelastic flow.\\
We first recall that within the contact area, the vertical displacement $u_z$ of the surface is given by
\begin{equation}
u_z=\delta-r^2/2R=\delta-(x^2+y^2)/2R
\end{equation}
In moving forward the sphere a small distance $\Delta$ along $x$ axis, the change in vertical displacement at $(x,y)$ is
\begin{equation}
\Delta \left( \partial u_z / \partial x \right)=-x\Delta/R
\end{equation}
so that the work done by the pore pressure over a strip of width $dx$ will be $dW=|p(x,y)dx(x\Delta/R)|$. This work is of a purely dissipative nature as it corresponds to the viscous dissipation involved in successive squeezing and re-swelling of the gel network at the leading and trailing edges of the contact, respectively. This is the reason why we take the absolute value of the product $p(x,y)x$ in the calculation.\\
The dissipated energy per unit advance of the sphere is by definition
\begin{equation}
\frac{W}{\Delta}=\frac{1}{R}\int\int_A \left |p(x,y) x \right |dxdy
\label{eq:work0}
\end{equation}
When $v<v_c$, i.e $\mathrm{Pe}<1$, $\zeta=1$ and $a=a_0$, the pore pressure $p(x,y)$ is
\begin{equation}
p=\frac{F_n}{2\pi a_0^2}\left ( 1-\frac{x^2+y^2}{a_0^2} \right )Pe \frac{r\cos \theta}{a_0}\:\:\:\:,\: r \leq a_0\:.
\end{equation}
thus
\begin{equation}
\begin{split}
\frac{W}{\Delta}&=\frac{F_n}{2\pi a_0^2} Pe \int_{-a_0}^{a_0}\int_{-\sqrt{a_0^2-y^2}}^{\sqrt{a_0^2-y^2}}  
\left | x^2 \left(1-\frac{x^2+y^2}{a_0^2}\right)\right| dxdy\\	
&=\frac{a_0}{6R}F_n Pe
\end{split}
\label{eq:Ft}
\end{equation}
The effective length of the crack front in our model is $\pi a_0$, so the energy dissipated per unit crack area $\mathcal{G}_{poro}$ is 
\begin{equation}
\mathcal{G}_{poro}\equiv \frac{W}{\Delta \pi a_0}=\frac{1}{6\pi}\frac {F_n}{R}Pe
\end{equation}
	%
%
\bibliographystyle{unsrt}

\end{document}